\documentclass[prl,twocolumn,letterpaper]{revtex4}

\usepackage{graphicx}
\usepackage{pslatex}
\usepackage{amsmath}

\newcommand{\beq}{\begin{equation}}
\newcommand{\eeq}{\end{equation}}

\begin{document}
\title{Stringent neutrino flux constraints on anti-quark nugget dark matter}

\author{
P.~W.~Gorham
}
\vspace{2mm}
\noindent
\affiliation{
Dept. of Physics and Astronomy, Univ. of Hawaii at Manoa, Honolulu, HI 96822. 
}
\author{
B. J. Rotter
}
\vspace{2mm}
\noindent
\affiliation{
Dept. of Physics and Astronomy, Univ. of Hawaii at Manoa, Honolulu, HI 96822. 
}

\begin{abstract}
Strongly-interacting matter in the form of nuggets of nuclear-density material
are not currently excluded as dark matter candidates in the ten gram to hundred kiloton 
mass range. A recent variation
on quark nugget dark matter models postulates that a first-order imbalance between
matter and antimatter in the quark-gluon plasma prior to hadron production in the early universe binds up
most of the dark matter into heavy (baryon number $B \sim 10^{25}$) anti-quark nuggets
in the current epoch, explaining both the dark matter preponderance and the matter-antimatter 
asymmetry. Interactions of these massive objects with normal matter in the Earth and Sun will
lead to annihilation and an associated neutrino flux in the $\sim 20-50$~MeV range. We
calculate these fluxes for anti-quark nuggets of sufficient number density to account for the dark matter
and find that current neutrino flux limits from Super-Kamiokande provide stringent constraints on several
possible scenarios for such objects. Conventional anti-quark nuggets in the previously allowed mass range 
cannot account for more than $\sim 1/5$ of the dark matter flux; if they are in a color-superconducting
phase, then their muon production during matter annihilation must be suppressed by an order of magnitude below 
prior estimates if they are to remain viable dark matter candidates.
\end{abstract}

\maketitle

Baryons in the form of normal matter are tightly constrained as dark matter (DM) candidates under
standard models of light element synthesis the early universe~\cite{DM0}, but a hypothesis originally
proposed by Witten~\cite{Witten84} provides for production of stable composite strange quark matter,
also known as  {\it nuclearites} or {\it strangelets}, 
prior to nucleosynthesis, effectively removing this baryonic material from interaction
during later phases~\cite{Itoh70}. In the decades since this proposal, 
it has undergone much scrutiny and refinement~\cite{Wil98,LH04},
and still remains as a viable hypothesis~\cite{SQNreview}. With the tightening constraints on beyond-standard model
particle DM candidates, alternative scenarios such as stable quark nuggets also deserve renewed attention. 
Although such objects might seem to be excluded 
as DM candidates because
of their strong interactions, it is in fact the cross section per unit mass of candidate material
$\sigma/M$ that is astrophysically important for the viability of DM candidates. 
The current best estimate of this constraint is 
$\sigma/M \lesssim 0.1-1$~cm$^2$~g$^{-1}$~\cite{CSS08,Peter13}.
This bound is easily evaded by high mass quark nuggets because their mean time per interaction
for direct detection exceeds the available observation time.

Mass regions from baryon number $B\sim 10^3-10^{25}$ are already 
largely constrained~\cite{Mack07,Price84,Price86,Zhit03}
to values well below the DM flux. Above this range
indirect detection methods must be employed; limits on seismic events
in the Earth and Moon is one example. In most scenarios, quark nuggets of normal matter,
rather than antimatter, are considered, and their interactions with other massive bodies
are primarily through collisional heating via their kinetic energy, since their mean 
velocity in the solar neighborhood should be of order 250~km~s$^{-1}$, consistent
with the galactic virial dispersion.

Because the nuggets are formed at a very early epoch, most likely prior to the origin of
the matter-antimatter asymmetry, both normal- and
anti-quark matter are allowed; in fact a nearly equal mix is possible. A moderate asymmetry  
in these objects during their
production and subsequent ``drop-out'' from normal matter interactions could naturally
explain both the DM and the current matter-anti-matter asymmetry, without a need for
fine-tuning~\cite{Zhit03,Oaknin05,FZ08}. 
This scenario results in a large population of anti-quark nuggets (AQN) as
well as normal quark nuggets (QN), and the former objects would have very different 
phenomenology of their interactions with normal matter, with antimatter annihilation
dominating over kinetic energy deposition. In this report, we consider a
proposed model where the DM consists of AQN and QN in the ratio
AQN:QN = 3:2~\cite{FZ08,FLZ10,Lawson11,Lawson12}. 
Matter-antimatter annihilation from those AQN entering 
the Earth and Sun lead to a detectable MeV neutrino flux from pion decay over a very wide
range of AQN masses. The current bounds from SuperK~\cite{SuperK08} then provide 
constraints that, under standard parameters for these objects, exclude them as dark
matter candidates.

To model AQN interactions in the Earth and Sun, we developed a Monte Carlo simulation which starts
with an isotropic flux of AQN at some distance from the target body, using a Maxwellian velocity distribution:
\begin{equation}
f(v) dv = \frac{4}{\sqrt{\pi}} \left ( \frac{3}{2} \right )^{3/2} \frac{v^2}{~v_{rms}^3}
~\exp{\left ( \frac{-3v^2}{~~2v_{rms}^2}\right ) dv}
\end{equation}
where $v$ is the AQN speed, and $v_{rms}\sim 270$~km~s$^{-1}$ is the galactic velocity dispersion.
For the Sun, the outer boundary was placed at 30AU; at this distance 
the solar escape velocity is about 5~km~s$^{-1}$, around 2\% of the mean speed of
the particles; thus the phase space is only slightly biased by the solar potential. 
For the Earth, we found that $10$ Earth radii was an adequate distance under similar considerations.
The relevant average inward flux of DM particles through 
a spherical surface, based on the assumed average local
DM density $\bar{\rho}_{DM} = 0.3$~GeV~cm$^{-3}$, is given
by $\Phi = \bar{n} \bar{v} / 2$, where $\bar{n} = \bar{\rho}_{DM}/m$ for AQN mass $m$.

Once
their speed and impact parameter are known, 
the impact parameter for capture is given by
\begin{equation}
b_{cap} = R \sqrt{1+ \frac{2GM}{R v^2}}
\end{equation}
where $G$ is the gravitational constant, and $R,M$ are the target body
radius and mass, respectively.
Those AQN that will intersect the capture disk are then propagated into the target body 
using a 4th-order Runge-Kutta integrator. 

Upon entering the target body interior, AQN begin to collide with and annihilate with normal matter.
In the Sun, matter is encountered both through their infall speed and geometric cross section, and through the 
rapid rise in the thermal velocity of the solar interior, which drives material into the AQN. 
For the former accretion mechanism,
the AQN speed and cross section are used, and for the latter, the acoustic speed (a close approximation for 
the thermal velocity) vs. solar
radius, combined with the area of the (presumed) spherical AQN are used. 
The kinetic energy loss by ram pressure for QN was first
derived in the analysis of de Rujula \& Glashow~\cite{Seis1}:
\beq
\label{kinetic}
\frac{dE}{ds} ~=~ -\sigma_n \rho(s) v^2
\eeq
where $\sigma_n$ is the cross sectional area of the nugget, and $\rho (s)$ is the density along the track $s$.
For normal quark matter nuggets, the mass remains constant throughout the interaction, but for AQN, 
matter annihilation at some fractional efficiency $\epsilon_{a}$
causes it to lose mass during its transit of the target body interior. The efficiency factor is
due to reflection of incoming nuclei by the surface of the AQN;
typical estimates of the efficiency are $\epsilon_{a} \sim 0.05$. Thus the complete
equation of motion for AQN is modified both by the mass loss and the efficiency.
\beq
m(t) \frac{d\vec{v}}{dt} = -(1-\epsilon_{a})\sigma_n(m) \rho(r) v^2 \hat{v} - \frac{G m(t) M_{int}(r)}{|r^3|}~\vec{r}
\eeq
where $M_{int}(r)$ is the target body mass interior to radial distance $r$.  The  variable AQN mass
term $m(t)$ appears in this form here since we have assumed that the energy
of annihilation is radiated isotropically from the AQN surface, a reasonable assumption.
The cross sectional area of the nugget is given by
\beq
\sigma_n(m) = \pi  \left ( \frac{3m(t)}{4\pi\rho_N} \right )^{2/3}
\eeq
where $\rho_N = 3.5 \times 10^{17}$~kg~m$^{-3}$ is the nuclear density of the nuggets.
The mass loss function $m(t)$ is determined by
\beq
\frac{dm}{dt} = \frac{dm}{ds}\frac{ds}{dt} = v \frac{dm}{ds} = F(v, c_s)
\eeq
where the function $F(v, c_s)$ accounts for the two accretion regimes with
respect to the solar acoustic velocity $c_s(r)$.
For $v> c_s$, the speed of the AQN is supersonic and cross-sectional capture
of material dominates; for subsonic speeds $v < c_s$, the 
capture becomes spherical, dominated by
the acoustic speed of the surrounding gas:
$$F(v>c_s) \simeq v \sigma_n \rho(r),~~~F(v<c_s) \simeq c_s(r) A_n \rho(r) $$
where $A_n = 4\sigma_n$ is the AQN total area.
For the solar acoustic speed and density, we use standard solar model data~\cite{stdsolar}.
With this equation of motion and mass loss function, the total number of nuclear interactions per
AQN is determined. For the Earth, a piecewise continuous model of the interior density was used.

Recent calculations by two independent groups~\cite{RSB,Bernal2013} 
of annihilation of weakly interactive massive particle (WIMP) dark
matter candidates have looked specifically at the neutrino production through WIMP
annihilation in the Sun. These calculations involved detailed simulations of the 
hadronic interactions in the high-density solar core, with a specific goal of yielding
the number of neutrinos produced per annihilation. While these calculations included
some uncertainty in the efficiency of producing final hadronic states from WIMP
annihilation, they otherwise apply directly to annihilation of AQN in the sun, a process
in which hadronic final states are dominant. While the average density of
the Earth is only about 4\% of the solar core, these estimates found very little dependence on density in their
estimates of number of pions produced. We assume these results apply to
the Earth as well, as supported by collider studies of 
low-energy $p\bar{p}$ annihilation~\cite{PDG}.

For the average number $N_{\nu}$ of
neutrinos per annihilation, reference~\cite{RSB} found $N_{\nu} \sim 1-10$, depending on details of
the hadronic fraction of the annihilation products. While $\pi^-$ form Coulombic
atoms and are eventually absorbed, the positive pions decay leading to three
neutrinos: $\pi^+ \rightarrow \mu^+\nu_{\mu} \rightarrow e^+\nu_e\bar{\nu}_{\mu}\nu_{\mu}$,
with energies up to about 53~MeV. We consider two cases of AQN structure. If
the AQN is standard strange matter, in a non-color-superconducting phase, the annihilations will produce
pions which will decay as in normal hadronic annihilations. In this case, for the neutrino yield for a
given AQN mass will be one neutrino per nucleon annihilated, a conservative value at 
the low end of the estimated range.

However, there are reasons to believe that
AQN might exist in a color-superconducting (CS) phase~\cite{Forbes:2006ba}; as such, 
pion production and decay via annihilations of nucleons within the AQN may be suppressed. 
For this case, we follow the analysis
of Lawson (2011)~\cite{Lawson11}, which estimated of order one emitted muon per nucleon
annihilated in CS-AQN; this was considered an upper limit for the muon emission rate. 
In this case, half of the resulting neutrinos
are $\bar{\nu}_{\mu}$ compared to 1/3 $\bar{\nu}_{\mu}$ in the normal hadronic annihilation case,
yielding a bound that is 50\% lower per nucleon annihilated.
Thus for the color superconducting case, no more than
$0.1$ muon per nucleon annihilated is allowed for
CS-AQN to remain a viable dark matter candidate, an order of magnitude below
the rate used by reference~\cite{Lawson11}.

Underground neutrino detectors such as
SuperK have their highest signal-to-noise ratio in the energy range from
20-50~MeV for $\bar{\nu}_e$ and $\nu_e$. 
Matter-enhanced oscillations then
play an important role in determining the neutrino flux observed at Earth.
From the solar core, the matter-enhanced
oscillation probability $P(\bar{\nu}_{\mu}\rightarrow \bar{\nu}_e) \simeq 1/6$,
with the resonance region extending out to about half the solar radius for neutrinos
in the 10-50 MeV energy range. 

We simulate the AQN mass range from $10^{-7}$ to $10^8$~kg. For reference,
at the high end of the range, the AQN have roughly the mass of an aircraft carrier,
and the diameter of the period at the end of this sentence. 
Beyond $10^8$~kg,
we find that the interval between individual AQN captures by the Sun begins to 
significantly exceed the 
time it takes for an AQN of this mass to deposit its energy in the Sun. Since the
neutrino emission is no longer continuous in this case, the Sun is effectively
too small to probe mass ranges beyond $10^8$~kg. For the Earth, a similar limit
obtains at AQN masses exceeding 1~kg. 

\begin{figure}[htb!]
\includegraphics[width=3.25in]{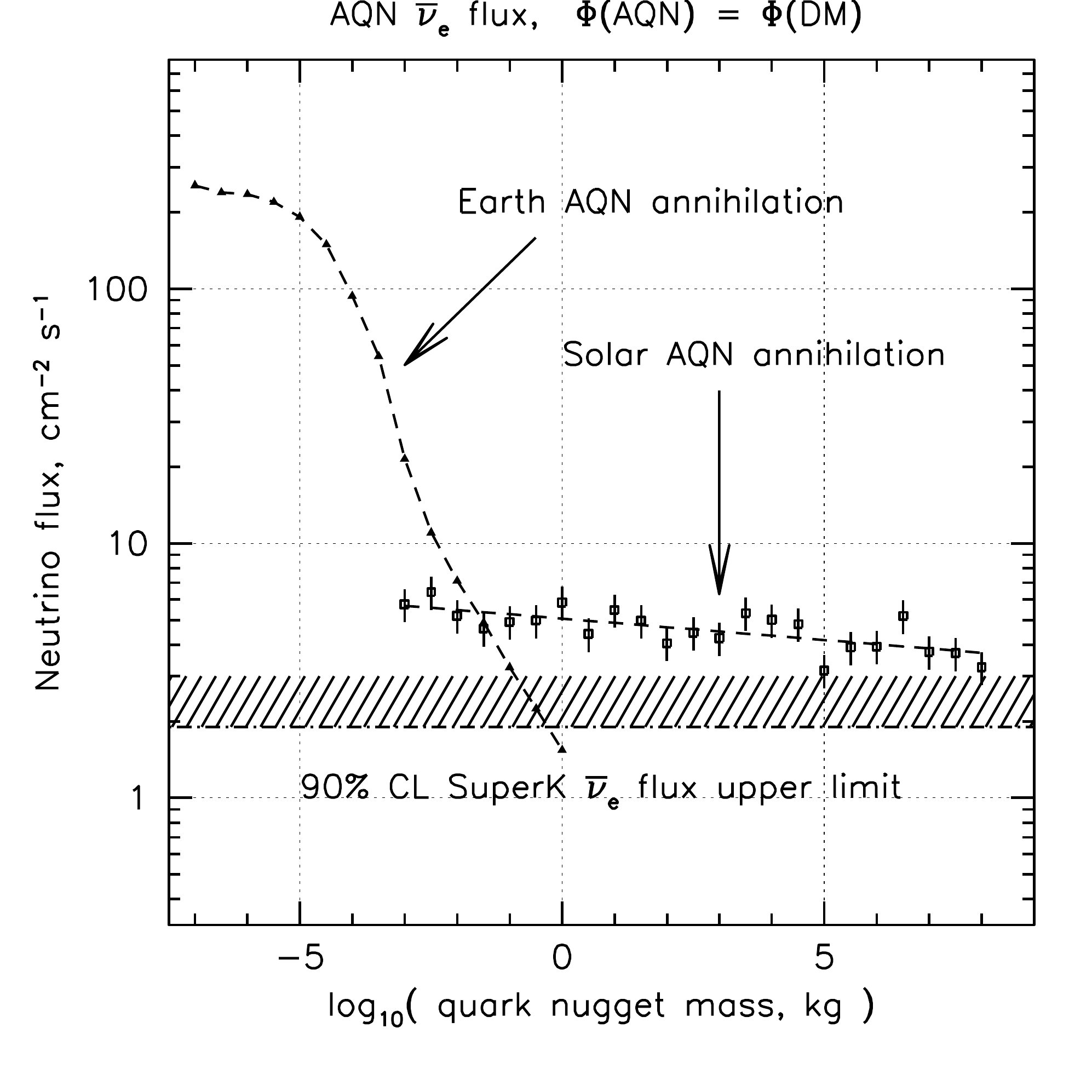}
\caption{$\bar{\nu}_e$ fluxes from the Sun resulting from AQN annihilation when the
AQN+QN flux equals the local DM flux for $\rho_{DM} = 0.3$~GeV~cm$^{-3}$.
Results to the left arise from AQN interactions in Earth, and to the right
in the Sun. Limits from SuperK for the $>~20$~MeV $\bar{\nu}_e$ flux 
are also shown. 
\label{nufluxes}}
\end{figure}

For our solar simulations we found that for all AQN masses above
1 gram, of order 90\% or more of the annihilation occurred within $R/R_{\odot} < 0.5$. 
Combining the production fraction and oscillation probabilities, 
we expect a fraction of 1/18 of the total solar AQN neutrino flux to appear in 
$\bar{nu}_e$ at Earth~\cite{RSB}.
Despite its small fraction of the total, the $\bar{\nu}_e$ component of the 
neutrino flux  provides the most stringent
limit for the solar AQNs, because of the high sensitivity of terrestrial neutrino detectors to 
this flavor channel.
For AQN neutrinos produced in the Earth, the matter oscillation resonance energies are in the 1-10~GeV range 
and there is thus no matter enhancement. The average vacuum oscillation probability gives
$P(\bar{\nu}_{\mu}\rightarrow \bar{\nu}_e) \simeq 1.3 \times 10^{-3}$, giving
an overall fraction of $\bar{\nu}_e / \nu_{total} \sim 4.3 \times 10^{-4}$ for 
Earth. Again, because of the high $\bar{\nu}_e$ sensitivity, this channel still provides
the best limits.

Results for the $\bar{\nu}_e$ flux at Earth's surface from both Solar and Earth
AQN annihilation are shown in Fig.~\ref{nufluxes}.
SuperK $\bar{nu}_e$ flux limits~\cite{SuperK08} are plotted along with the expected
all-neutrino and $\bar{nu}_e$ fluxes for an AQN+QN flux equal to the
expected DM flux. The resulting $\bar{nu}_e$ fluxes exceed the
limit by two orders of magnitude at the low-mass end from Earth fluxes; a factor of six where the Solar
limits take over; and 
and a factor of four at the high end of the mass range from Solar constraints. 
For the terrestrial AQN interactions, the large increase and plateau at lower masses
is due to an increase and eventual saturation in the capture probability of AQNs in the Earth.
For the solar case virtually all AQNs are captured and completely annihilated. 
The neutrino fluxes from AQN
interactions decrease slowly with AQN mass, primarily due to the increased
probability that AQNs of larger masses may cross the Sun on an unbound transit
chord, and survive to escape without depositing all of their mass-energy.

\begin{figure}[htb!]
\includegraphics[width=3.15in]{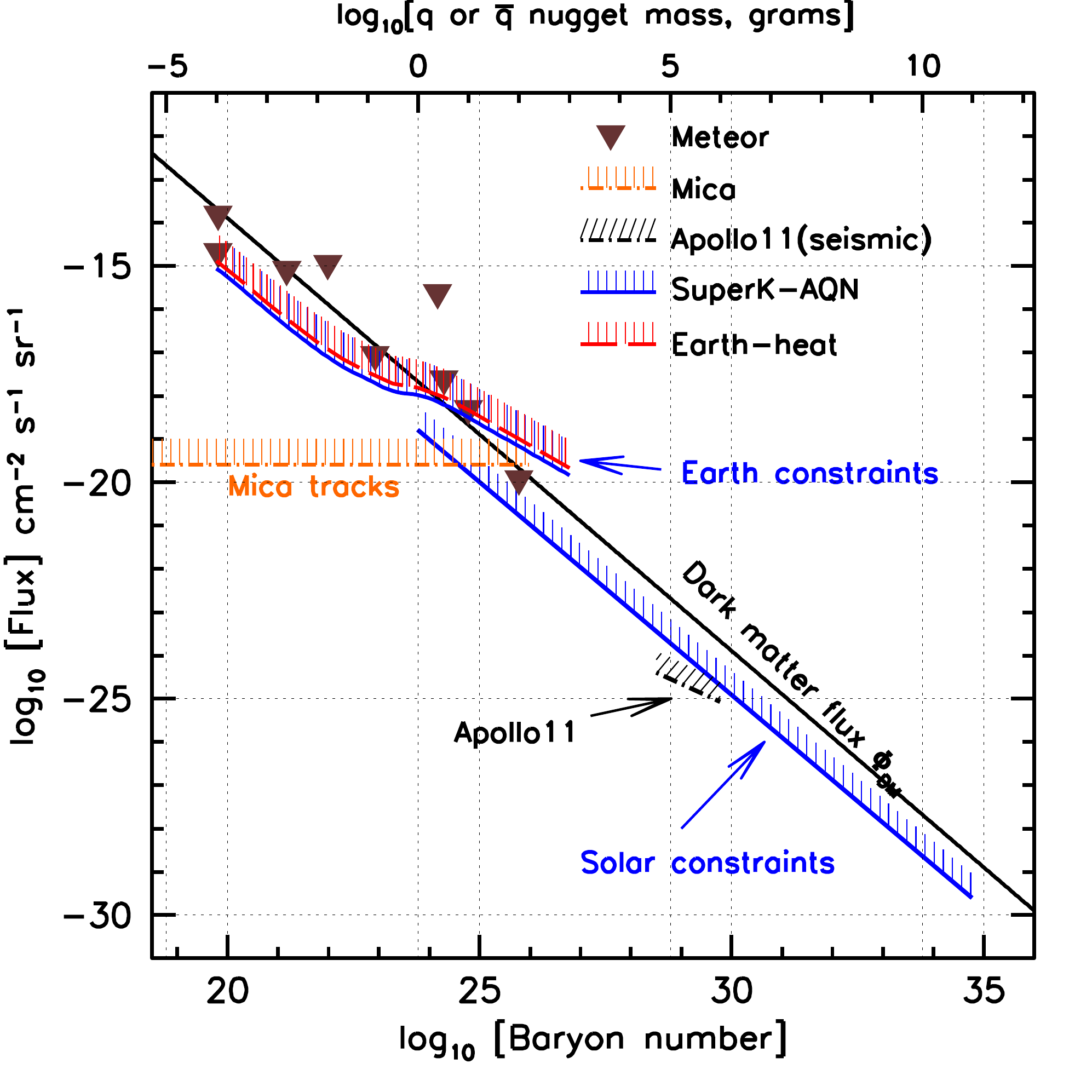}
\caption{Constraints on QN from prior work as
shown, along with neutrino and Earth heat flux
constraints from this work on AQN.
\label{qnlim}}
\end{figure}

Since the results for Fig.~\ref{nufluxes} are computed for an assumed flux of
AQNs equal to 3/5 of the DM flux at a mean 3-dimensional velocity of 
250~km~s$^{-1}$ and a density of 0.3~GeV~cm$^{-3}$, the Super-K limits from
$\bar{nu}_e$ translate directly to mono-mass limits on their fraction 
of the total DM flux as shown in Fig.~\ref{qnlim}.  
The plot shows also the equivalent differential DM flux (AQN+QN) for these masses, 
along with constraints
on QN which apply also to AQN as well in this case.
The curves in each case (and the data points where shown) are mono-mass
limits, treating the QN spectrum as a delta-function in mass. 
Integral limits involving (perhaps) more realistic spectra with a distribution of
masses would of course be more restrictive, and the limits shown are thus
conservative. The DM flux is thus also expressed as a pure differential
flux density of mono-mass objects at each given baryon number.

Prior limits (typically at the 90\% CL where stated) from the 
absence of observations of fast meteors~\cite{Porter85,Porter88}, appropriate
tracks in samples of the mineral Mica~\cite{Price86}, and seismic limits from the Apollo-11 
Lunar lander~\cite{Seis3} already provide constraints below the DM flux for all QN
of baryon numbers from $\sim 10^{19-25}$ from the terrestrial observations, 
and for a smaller window from 50-1000~kg based on the Lunar seismic limits.
Our results eliminate AQN as predominant DM 
over a wide mass range from $B=10^{20-35}$,
with no more than 15-25\% of the DM flux allowed in AQN over a wide
range where there were no previous constraints.
Our limits do not of course work in reverse for normal-matter QN; 
their phenomenology does not include neutrino production in the same manner as AQNs.

The typical total annihilation power produced in the Sun by AQNs of all masses 
at the 3/5 DM flux level is of order $10^{27}$~ergs~s$^{-1}$, less than
$10^{-6}$ of the solar luminosity; thus the energy deposited by AQN in the Sun
is of no consequence in constraining their flux. For the Earth, the results
are quite different: for AQN masses below about 1~g, the increasing capture rate of the
DM flux leads to heat generation in
the Earth that exceeds 10~TW, which is about the maximum allowed excess heat
that is not accounted for by other processes. Terrestrial power limits on
AQN fluxes are estimated in reference~\cite{Gorham12}, and we update that limit
here with a more precise estimate. In \cite{Gorham12}, the constraint
was given as a general exclusion of all AQN fluxes below $B \sim 10^{24}$; with
our simulation, we estimate that the maximal allowed AQN fluxes in the range
of $B \sim 10^{20-23}$ are of order 5-50\% of the DM flux, as plotted
in Fig.~\ref{qnlim}. However, the neutrino constraints are about a factor of 
three stronger than that of the terrestrial heat flux.

Our solar-based limits have virtually no dependence on the 
assumed 5\% AQN-nucleon capture efficiency,
at least within an order of magnitude.
This is because the great majority of deposited energy by AQN in the Sun comes
from objects whose orbits rapidly decay in the interior of the Sun after
they enter it, whether or not they were tightly bound to begin with. 
Capture efficiencies even an order of magnitude lower than the expected 5\% result
in a redistribution of the deposited energy, but the bound remains essentially unaffected.
At the lower masses, most of the AQN are found to fully annihilate outside
the solar core; lower nucleon capture efficiencies result in their
energy being deposited closer to the core, but do not allow them to escape.
The expected capture efficiency would have to be several orders of magnitude
below expected values~\cite{FZ08} in order to relax these bounds.
For the Earth-based limits, a large decrease in annihilation efficiency would
shift the plateau of the neutrino curve to lower energies, and would impact
the behavior of the limit, but since this regime is already covered by both
the Mica-based limits, and the terrestrial power flux constraints, it 
is unimportant to the main results here.

In summary, while AQN of masses in the 1~kg to 100~Kton range
may be stable and could still provide a component of the current closure
density of the universe, their interactions in the Sun are a strong source of
neutrino fluxes at Earth, and preclude their number density to a fraction
of no more than 25\% of the DM density. For $\rho_{DM}$ we have
assumed a value at the low end of current estimates~\cite{DM1,DM2}; values 
could be several times higher than our estimate, and would lower the allowed
AQN fraction of the DM accordingly. Our results are also robust to
assumptions regarding the annihilation efficiency of AQN; those entering 
the Sun are found to be unlikely to escape complete destruction even 
at efficiencies an order of magnitude below expectations. It is
rather striking that while the excess of deposited energy by a dark
matter flux of AQNs is found to have negligible effects on
the solar luminosity, neutrino fluxes from the Sun are remarkably
sensitive to these objects, yielding robust and stringent constraints
over an extensive, previously-allowed mass range. Finally,
AQN in a color-superconductive phase may still evade the bound, but only
if their neutrino production is greatly suppressed in either
number or neutrino energy compared to current expectations. 

We thank J. Learned and J. Kumar for very useful discussion, 
and K. Lawson and A. Zhitnitsky for comments on the manuscript.
We are grateful to the US Department of Energy, High Energy Physics Division,
and NASA for their generous support of this research.

\end{document}